# Whether the Health Care Practices For the Patients With Comorbidities Have Changed After the Outbreak of COVID-19; Big Data Public Sentiment Analysis

Bilal Ahmad[a,b], Sun Jun[a,d]

[a]*Department of Internet of Things, Jiangnan University, Wuxi, China*
[b]*bilalrouf@yahoo.com*, [c]*bilalrouf@gmail.com*
[d]*sunjun_wx@hotmail.com*

## Abstract

After the pandemic of SARS-CoV-2, it has influenced health care practices of all the world. Initial investigations indicate that patients with comorbidities are more fragile to this SARS-CoV-2 infection. They suggested postponing the routine treatment of cancer patients. However, few meta-analyses suggested evidence are not sufficient to hold the claim of the frailty of cancer patients to COVID-19 and they are not in the favour of shelving the scheduled procedures. There are recent studies in which medical professionals, according to their competence, are referring to change the routine practices on how to manage the applicable therapeutic resources judiciously to combat against this viral infection. This is a different study which reveals the cancer patients' viewpoint about how health care practices have been changed in their opinion? Are they satisfied with their treatment or not? To serve the purpose, we gathered more than 60000 relevant tweets from twitter to analyse the sentiment of cancer patients around the world. Our findings demonstrate that there is a surge in the argument about cancer and its treatment after the outbreak of COVID-19. Most of the tweets are reasonable (52.6%) compared to negative ones (24.3%). We developed polarity and subjectivity distribution to better recognize the positivity/negativity in the sentiment. which reveals that the polarity range of positive tweets is within the range of 0 to 0.5. which means the tendency in the tweets is not so much positive. It is a modest statistical evidence in the support of how natural language processing (NLP) can be accepted to better understand the patient's behaviour in real-time and it may facilitate the medical professionals to make better decision to organize the routine management of cancer patients.

## *Introduction*

This pandemic situation has set the stage for artificial intelligence, machine learning and natural language processing (NLP) tools to assist the health care professionals to prove their worth as a supporting tool. Unstructured data is increasing with an immense speed around the globe on the internet. NLP has been used to extract useful information from the growing amount of data automatically. Primary research related to sentiment analysis was to understand the consumer's feedback towards services they were being offered. It was mostly a business strategy in the past, it is now used in every field of life that carries from landscape management of community to medical resources administration. Following are few studies in various fields which have used natural language processing (NLP).

Richad A. discovered the effect of gardens on community sentiment. He built up a quantitive measure that characterizes the individuals' positive/negative point of view while they are in and outside the park. He found that people in some areas of New York tweets positively while they are inside the park as compared to when they are outside the park (Plunz et al., 2019). During the outbreak of pandemic H1N1, Chew C. used the twitter data to classify the tweets into sarcasm and humour. The objectives of the study included validating whether Twitter can be adopted as real-time sentiment analyzer or not (Chew & Eysenbach, 2010). Sunir G. used the sentiment analysis of patients feedback and with the help of machine learning algorithms, they tried to predict how comfortable patients' visit to the hospital was? Whether they treated well the patient and if he would recommend this hospital to others? (Gohil et al., 2018)

Recent studies are stating that patients with comorbidities are frailer to the COVID-19 and clinical practice for those patients should be changed to reduce the risk of being infected with SARS-CoV-2(Landman et al., 2020). Some researchers showed some concerns that shreds of evidence are not enough to stat that patients with comorbidities are more fragile to SARS-CoV-2(H. Wang & Zhang, 2020). Later on, metadata analysis in different countries suggested that there are few, not all, comorbidities such as diabetes, hypertension, cardiovascular and COPD are most common in COVID-19 patients and medical experts have suggested various practices they are observing to handle their routine patients(Richardson et al., 2020; B. Wang et al., 2020).

All the research till now presents the medical experts' perspective about COVID-19 patients with comorbidities. We don't find; to the best of our knowledge, any study addressing the issue in patients' perspective. What do they think how frequent their regular appointments for the treatment are cancelling or rescheduling during the pandemic days? Whether the regular standard procedures for their treatment has been changed in the view of cancer patients and the people who are associated with them? We tried to get the statistical evidence based on available data to address the above-mentioned questions with the help of NLP. To obtain the public perspective we did sentiment analysis of twitter data which includes tweets more than 60000.

## *Design & Data Collection*

We used Jupyter notebook as code editor and code is written in the python programming language. Ethical approval was waived off owing to the anonymity of data. We started collecting tweets dataset related to comorbidities from 25th to 30th April. We used the "Twitterscraper" tool in the command line to retrieve the 61400 tweets related to coronavirus and cancer published in the range of January 1, 2020, to April 25, 2020. We started writing our code on 1$^{st}$ May 2020. We made the following 15 search queries to search the relevant tweets.
1.'Covid-19 and cancer' 2.'Coronavirus and cancer' 3.'Covid-19 and oncology' 4.'Coronavirus and oncology' 5.'Cancer and surgery' 6.'Cancer and radiotherapy' 7.'Coronavirus and radiotherapy' 8.'Coronavirus and cancer treatment' 9.'Cancer treatment appointment' 10.'Chronic diseases and COVID-19' 11.'Chronic disease and coronavirus' 12.'Coronavirus and cancer surgery' 13.'Chemotherapy and covid-19' 14.'Chemotherapy appointment cancelled' 15.'Radiotherapy appointment cancelled'. Queries are not case sensitive. Twitterscraper, unlike many other APIs, has the advantage to retrieve the tweets which are older than 7 days(*TwitterScraper*, n.d.).

## *Data Processing*

At this stage, hashtags, meaningless words, special characters, retweets and duplicate tweets were removed. We got 41642 tweets after cleaning the dataset. The number of tweets published each month within the range of search terms indicates that there is an increase in discussion among public related to comorbidities & cancer after the outbreak of COVID-19. The phenomenon can be observed in figure 1 which presents the number of tweets published each week including all search terms, after removing the duplicate and retweets.

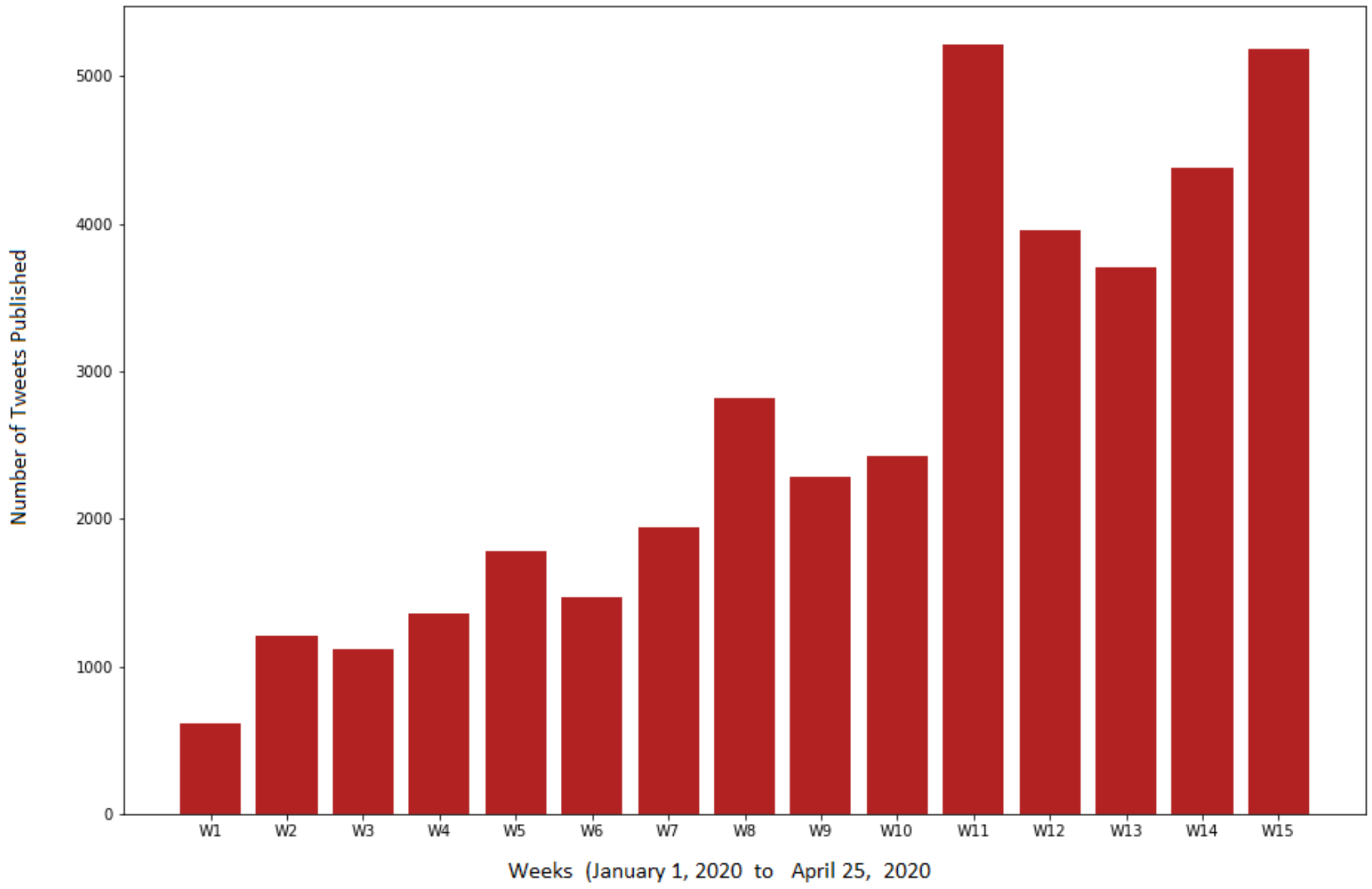

Figure 1. Number of tweets published per week from January 1, 2020, to April 25, 2020

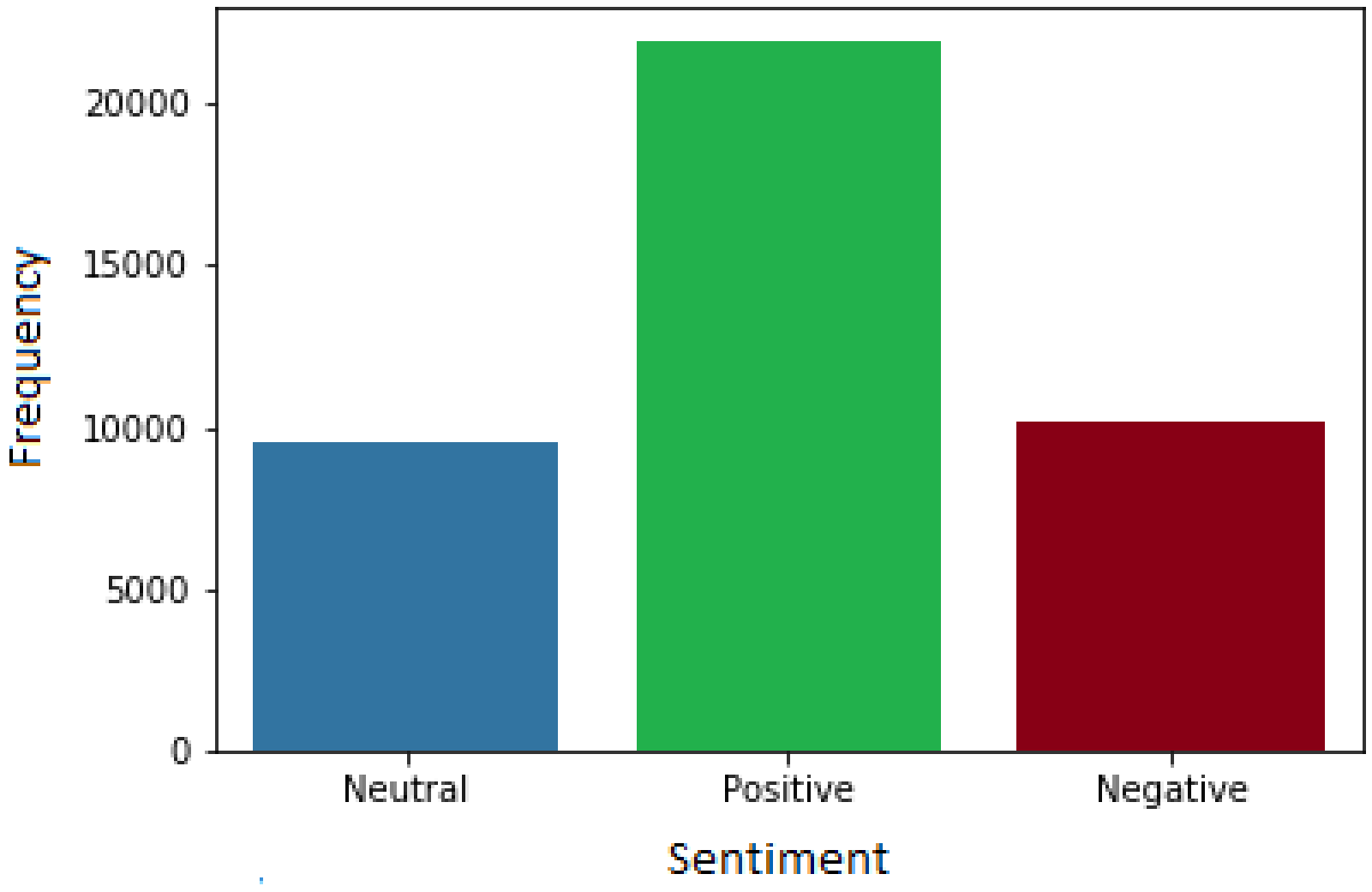

Figure 2. Number of positive, negative and neutral Tweets after the sentiment analysis. Positive = 21928; Negative= 10134; Neutral= 9580

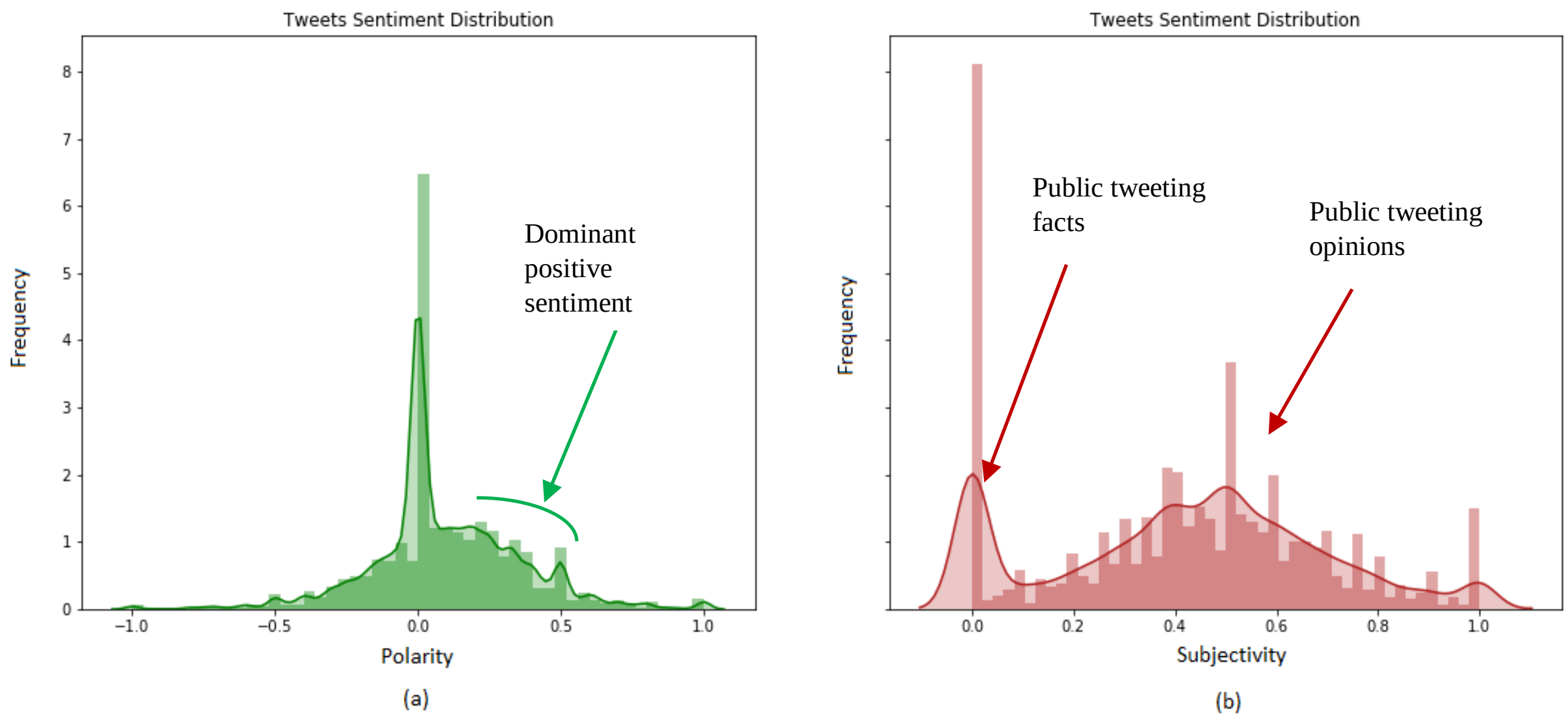

Figure 3. Polarity and subjectivity distribution of tweets.

## *Sentiment Analysis & Discussion*

The measure of any kind of opinion & fact in the text is called subjectivity. Its scope varies from 0 to 1. The value near 1 indicates the text is more likely an opinion, and near to 0 shows a fact. Polarity is the measure of emotion in the text. The basic emotions are either positive, negative or neutral and are measured in the range of -1 to 1 quantitively, where 0 indicates a neutral emotion. To measure the subjectivity & polarity quantitively, we used the TextBlob python library. Here is a basic description of how does TextBlob works. In this library each word has given some id called "WordNet ID" and it gives each WordNet Id a different polarity and subjectivity. A single word may have different quantitative polarity & subjectivity value according to the sense it conveys. A word "great", for example, is assigned the values of 0.4 & 0.2 for polarity and subjectivity respectively if it conveys the sense of large in size or big in number. The same word "great", however, would be assigned the values of polarity & subjectivity equal to 0.8 each and different WordNet id if it reveals the perceiving of remarkable, extraordinary or greater in effect.(TextBlon, n.d.)

Figure 2 illustrates that most numbers of retrieved tweets were positive (21928 out of 41642). Negative & neutral tweets were approximately equal in number (negative=10134, neutral=9580). Any value of polarity higher than 0 is weighed as positive sentiment and in this situation, most of the polarity values lie between 0 to 0.5. Polarity and subjectivity distribution in figure 3, produces the further precise picture of sentiment. It is apparent from figure 3(a) that polarity of tweets ranges from -0.5 to 0.5. Positive sentiment has bit higher amplitude, a spike can be observed at the pointed out the region of figure 3(a) in the range of 0. 0.5. We can infer from this narrow range of polarity people are tweeting positively about their treatment. However, it is bear a minimum positive range of sentiment. Figure 3(b) indicates there is a range of opinion (0.2-0.8) in the tweets of public including some facts as well. A sharp spike at 0 in figure 3(b)indicates public perhaps would be stating facts about COVID-19 and higher amplitude at 0.5 indicates people are expressing their emotions and opinions. Both points are highlighted in figure 3(b).
Figure 4 represents the cloud words of all the 60000+ tweets. Larger the font size, greater the frequency of the word in the tweets. The most frequent words in the word cloud are the search queries which were used to retrieve the tweets and most common discussion topic words. The word cloud indicates that people are speaking more about breast and lung cancer, although breast cancer and lung cancer were never in the search terms. It makes sense, as one of them is the most common type of cancer and the other one is directly linked with the symptoms of COVID-19 patients.
It would suggest people are worried about their regular treatment or rescheduling of their routine visits If the world cloud figure would have contained words such as alarmed, worried, delayed, postponed or cancelled etc. Increase in the number of tweets related to comorbidities (figure 1), most positive number of tweets (figure 2), polarity & subjectivity distribution analysis (figure 3) and word cloud image (figure 4), all these four supports the statements that there is an increase in discussion among the public from January to April 25, 2020, about the comorbidities and COVID-19. But cancer patients or the individuals associated with them don't think their regular treatment is affected by COVID-19. Even if it is affected, they are not talking about it negatively on public forums and social media platforms.

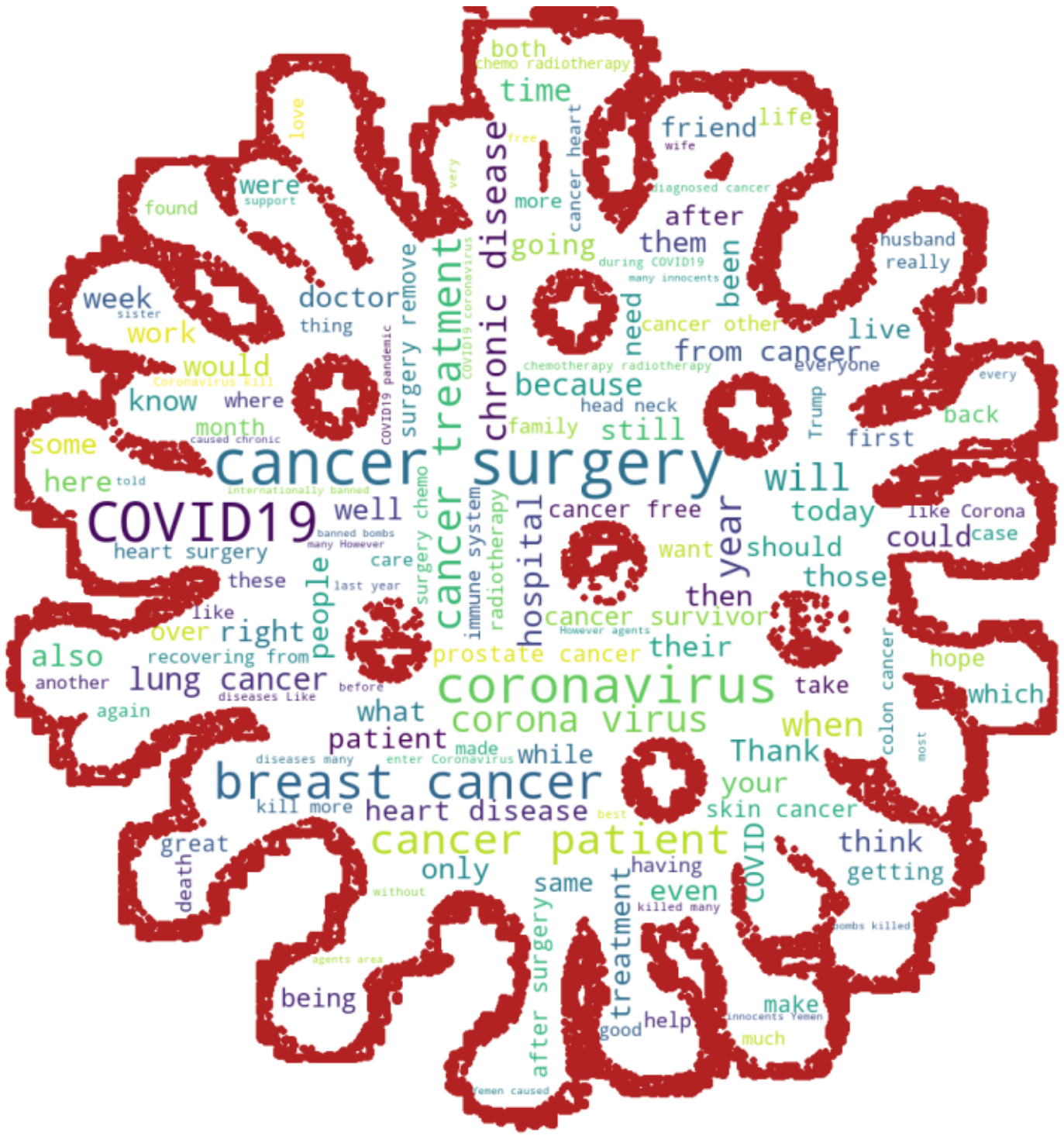

Figure 4. Word Cloud image from the tweets dataset.

## *Limitations & Future Aspects of Study*

The tool, "Twitterscraper", we used in our investigation to elicit the tweets has the advantage (able to fetch tweets older than 7 days) over several APIs but it got some limitations as well. It doesn't fetch all the tweets published & associated with searched terms. Infect, Twitter doesn't support any API to squeeze all the tweets for any search query. Moreover, the privacy settings of twitter users also play an important role in whether to permit an API to retrieve the tweets. So the dataset is incomplete in a sense it doesn't have all the tweets published by public related to comorbidities.
As investigation's results reflect the positive sentiment of public about their treatment, we tried to be more skeptical owing to the fact dataset was incomplete and to reduce the bias if there was any in the dataset, we included the inquiry queries such as "Radiotherapy appointment cancelled" and "Chemotherapy appointment cancelled”, which indicates the negative sentiment. Although, the privacy settings and probability of less retrieval of data would affect the tweets with positive and negative sentiment, equally.
Machine learning and deep learning algorithms can be trained to classify the sentiment more than 3 basic classes, to serve the purpose only labelled data is required to feed these networks.

With due code of ethics and the consent of patients, medical experts may observe patients’ behaviour to understand the effect of specific medication in real-time with the help of social media & NLP. Instead of general observation of patients, the experts can extend this practice to various geographic regional levels to relate to the other communal, economical, social factors in human behaviour. Standard working procedures would be required to develop to protect patient privacy and to get permission from social media firms to access public content for research purpose.

## *Conclusion*

After the outbreak of COVID-19, there is an increase in the discussion on social media about patients with comorbidities. 52% positive sentiment tweets (figure 2), their further polarity & subjectivity distribution (figure 3) and words cloud of tweets (figure 4) indicate that routine practices for cancer patients has not been affected very much in patients’ point of view. Even if their appointments are rescheduled or cancelled they are not expressing it on social media just as a positive gesture to the patients who are suffering from SARS-CoV-2. It is a modest statistical evidence in the support of how natural language processing (NLP) can be accepted to better understand the patient’s behaviour in real-time and it may facilitate the medical professionals to make a better decision to organize the routine management of cancer patients.

## Conflict of Interest

We declare no competing interests.

## References

Chew, C., & Eysenbach, G. (2010). Pandemics in the age of Twitter: Content analysis of tweets during the 2009 H1N1 outbreak. *PLoS ONE*, *5*(11), 1–13. https://doi.org/10.1371/journal.pone.0014118

Gohil, S., Vuik, S., & Darzi, A. (2018). Sentiment analysis of health care tweets: Review of the methods used. *Journal of Medical Internet Research*, *20*(4). https://doi.org/10.2196/publichealth.5789

Landman, A., Feetham, L., & Stuckey, D. (2020). *Cancer patients in SARS-CoV-2 infection : a nationwide analysis in China*. *2045*(20), 335–337. https://doi.org/10.1016/S1470-2045(20)30096-6

Plunz, R. A., Zhou, Y., Carrasco Vintimilla, M. I., Mckeown, K., Yu, T., Uguccioni, L., & Sutto, M. P. (2019). Twitter sentiment in New York City parks as measure of well-being. *Landscape and Urban Planning*, *189*(July 2018), 235–246. https://doi.org/10.1016/j.landurbplan.2019.04.024

Richardson, S., Hirsch, J. S., Narasimhan, M., Crawford, J. M., McGinn, T., Davidson, K. W., Barnaby, D. P., Becker, L. B., Chelico, J. D., Cohen, S. L., Cookingham, J., Coppa, K., Diefenbach, M. A., Dominello, A. J., Duer-Hefele, J., Falzon, L., Gitlin, J., Hajizadeh, N., Harvin, T. G., … Zanos, T. P. (2020). Presenting Characteristics,

Comorbidities, and Outcomes Among 5700 Patients Hospitalized With COVID-19 in the New York City Area. *Jama*, *10022*, 1–8. https://doi.org/10.1001/jama.2020.6775

TextBlon. (n.d.). *TextBlob Python Library*. https://textblob.readthedocs.io/en/dev/

*TwitterScraper*. (n.d.). https://github.com/taspinar/twitterscraper

Wang, B., Li, R., Lu, Z., & Huang, Y. (2020). Does comorbidity increase the risk of patients with covid-19: Evidence from meta-analysis. *Aging*, *12*(7), 6049–6057. https://doi.org/10.18632/AGING.103000

Wang, H., & Zhang, L. (2020). Risk of COVID-19 for patients with cancer. *The Lancet Oncology*, *21*(4), e181. https://doi.org/10.1016/S1470-2045(20)30149-2